# Constraints on the Scalar-Tensor theories of gravitation from Primordial Nucleosynthesis

A. Serna and J. M. Alimi

*DAEC, Observatoire de Paris-Meudon, 92195-Meudon, France*

(October 31, 1995)

## Abstract

We present a detailed calculation of the light element production in the framework of Scalar-Tensor theories of gravitation. The coupling function $\omega$ has been described by an appropriate form which reproduces all the possible asymptotic behaviors at early times of viable scalar-tensor cosmological models with a monotonic $\omega(\Phi)$. This form gives an exact representation for most of the particular theories proposed in the literature, but also a first-order approximation to many other theories. In most of scalar-tensor theories, the comparison of our results with current observations implies very strong bounds on the allowed deviation from General Relativity (GR). These bounds lead to cosmological models which do not significantly differ from the standard Friedman-Robertson-Walker ones. We have found however a particular class of scalar-tensor theories where the expansion rate of the universe during nucleosynthesis can be very different from that found in GR, while the present value of the coupling function $\omega$ is high enough to ensure compatibility with solar-system experiments. In the framework of this class of theories, right primordial yields of light elements can be obtained for a baryon density range much wider ($2.8 \lesssim \eta_{10} \lesssim 58.7$) than in GR. Consequently, the usual constraint on the baryon contribution to the density parameter of the universe can be drastically relaxed ($0.01 \lesssim \Omega_{b0} \lesssim 1.38$) by considering these gravity theories.

This is the first time that a scalar-tensor theory is found to be compatible both with primordial nucleosynthesis and solar-system experiments while implying cosmological models significantly different from the FRW ones.

# I. INTRODUCTION

From a theoretical point of view, the most natural alternatives to General Relativity (GR) are scalar-tensor theories which contain, in addition to the metric tensor, $g_{\mu\nu}$, a dynamical scalar field, $\phi$, the relative importance of which is determined by an arbitrary coupling function $\omega(\phi)$. In recent years, this class of metric theories has received a renewed interest [1,2], because it provides a natural (non-fine-tuned) way to restore the original ideas of inflation while avoiding the cosmological difficulties coming from the vacuum-dominated exponential expansion obtained in GR. Scalar-tensor theories also arise in current theoretical attempts at deepening the connection between gravitation and the other interactions. For example, in modern revivals of the Kaluza-Klein theory and in supersymmetric theories with extra dimensions, one or several scalar fields arise in the compactification of these extra dimensions [3–9]. Furthermore, scalar-tensor theories may also appear as a low-energy limit of superstring theories [10]

Several scalar-tensor theories have been proposed to date: 1) *Brans-Dicke's* theory [11], where $\omega$ = constant $\neq$ -3/2, 2) *Dirac's* theory [12], with $\omega$ = -3/2, 3) *Barker's* theory [13], where the gravitational coupling constant is effectively constant, 4) *Bekenstein's* theory [14], with variable rest mass, and 5) *Schmidt-Greiner-Heinz-Muller's* theory [15], which also includes a possible mass term for the scalar field. They are all particular cases of the general scalar-tensor theory by Bergmann, Wagoner and Nordtvedt [16–18]. In addition, the *scale-covariant* theory of Canuto [19,20] has a mathematical representation similar in many aspects to Dirac's gravitation but with a non-dynamical scalar function $\beta$.

The viability of a given alternative gravity theory can be analyzed by means of two kinds of tests [21]: those which examine its weak field limit and those which prove its full exact formulation. The first mainly consists of comparing the theory predictions in the limit of weak gravitational fields and slow motions with Post-Newtonian experiments. The only metric theory which is discarded by these experiments is Dirac's gravitation [21]. On the other hand, strong field tests consist of matching up the exact theory predictions to experiments. This is mainly achieved by means of cosmological models, whose predictions have to be consistent with any present observation of cosmological interest.

The astronomical data leading to the strongest bounds on the alternative gravity theories are the light element abundance observations, which have to be explained as an outcome of the primordial nucleosynthesis process (PNP). Some authors [22–26] have got a first insight into these PNP constraints by including a *constant* speed-up factor $\xi$ in the usual GR expression of the universe expansion rate. Such a simple approach implies rather stringent bounds on $\xi$. For example, Barrow [23] obtained $0.8 \leq \xi \leq 1.2$ while, using more recent reaction rates and observational data, Casas et al. [26] found an upper limit of 1.02.

Since, in a general scalar-tensor theory, $\xi$ is not necessarily constant at early times, the above approach only explores a very limited range of models. In principle, if $\xi$ varies during nucleosynthesis, the resulting bounds could be very different from those derived from the constant-$\xi$ approximation. Other authors [27–35] have thus preferred a more rigorous approach which consists of solving, from numerical computations, both the cosmological and nuclear equations. Since this approach follows the time evolution of $\omega(\phi)$ and all the dynamical functions, it has the additional advantage of determining how the PNP bounds result in present limits on the parameters. Using this approach, cosmological models and their corresponding light-element production have been analyzed in all the particular scalar-tensor theories proposed in the literature. The resulting bounds were always very stringent, implying that the only viable models are those whose predictions do not significantly differ from the standard GR ones up to at least temperatures of $10^{10}$ K.

In a previous paper [36] we found all the possible early behaviors of scalar-tensor cosmological models with a monotonic, but arbitrary, $\omega(\phi)$ function. These behaviors exhibited a variety of models much wider than that contained in all the previous particular cases. The aim of this paper is to perform a detailed numerical study of primordial nucleosynthesis bounds on theories presenting each one of these possible early behaviors. Such a study will allow us to elucidate whether the strong bounds previously found in some particular theories are also expected for any other one or, on the contrary, whether there exist some cases in which such bounds can be considerably relaxed. To that end, we will consider throughout this paper a standard scenario (i.e., a homogeneous and isotropic universe with vanishing cosmological constant, without exotic particles, etc.) in the framework of scalar-tensor theories.

The paper is arranged as follows. We begin outlining in Sect. II scalar-tensor theories and the basic equations to analyze the light-element production. We also introduce in that section a representation for the coupling function. Predicted abundances for a sweep of initial conditions and the constraints obtained from comparison with observations are shown in Sect. III. Finally, conclusions and a summary of our results are given in Sect. IV.

## II. SCALAR-TENSOR THEORIES

### A. Field equations and Cosmological Models

The most general action describing a massless scalar-tensor theory of gravitation is [16–18]

$$S = \frac{1}{16\pi} \int (\phi \mathcal{R} - \frac{\omega(\phi)}{\phi} \phi_{,\mu} \phi^{,\mu}) \sqrt{-g} d^4 x + S_M \quad (1)$$

where $\mathcal{R}$ is the curvature scalar of the metric $g_{\mu\nu}$, $g \equiv det(g_{\mu\nu})$, $\phi$ is the scalar field, and $\omega(\phi)$ is an arbitrary coupling function determining the relative importance of the scalar field.

The variation of Eq. (1) with respect to $g_{\mu\nu}$ and $\phi$ leads to the field equations:

$$\mathcal{R}_{\mu\nu} - \frac{1}{2} g_{\mu\nu} \mathcal{R} = -\frac{8\pi}{\phi} T_{\mu\nu} - \frac{\omega}{\phi^2} (\phi_{,\mu} \phi_{,\nu} - \frac{1}{2} g_{\mu\nu} \phi_{,\alpha} \phi^{,\alpha})$$
$$- \frac{1}{\phi} (\phi_{,\mu;\nu} - \Box \phi) \quad (2)$$

$$(3 + 2\omega) \Box \phi = 8\pi \phi - \omega' \phi_{,\alpha} \phi^{,\alpha}) \quad (3)$$

where $\omega'$ denotes $d\omega/d\phi$ and $\Box \phi \equiv g^{\mu\nu} \phi_{,\mu;\nu}$. In addition to Eqs. (2) and (3), we have the standard conservation law $T^{\mu\nu}_{;\nu} = 0$, where $T^{\mu\nu}$ is the energy-momentum tensor.

In order to build up scalar-tensor cosmological models, we consider a homogeneous and isotropic universe. The line-element has then the Robertson-Walker form:

$$ds^2 = -dt^2 + R^2(t)[\frac{dr^2}{1-Kr^2} + r^2 d\Omega^2] \tag{4}$$

and the energy-momentum tensor corresponds to that of a perfect fluid

$$T^{\mu\nu} = (\rho + P/c^2)u_\mu u_\nu + P g_{\mu\nu} \tag{5}$$

where $K = 0, \pm 1$, $R(t)$ is the scale factor, $\rho$ and $P$ are the energy-mass density and pressure, respectively, and $u_\mu$ is the 4-velocity of the fluid. The field equations (2)-(3) can be then written, in terms of $H \equiv \dot{R}/R$ and $D \equiv \dot{\phi}/\phi$, as

$$\dot{H} = -\frac{c^2 K}{2R^2} - \frac{3}{2}H^2 + \frac{1}{2}DH - \frac{4\pi}{(3+2\omega)\phi}(\rho - 3P/c^2)$$
$$-\frac{4\pi}{\phi}P/c^2 - \left(\frac{\omega}{4} - \frac{\omega'\phi}{2(3+2\omega)}\right)D^2 \tag{6a}$$

$$\dot{D} = -D^2 - 3DH + \frac{8\pi}{(3+2\omega)\phi}(\rho - 3P/c^2)$$
$$-\frac{\omega'\phi}{(3+2\omega)}D^2 \tag{6b}$$

where dots mean time derivatives.

By eliminating $\ddot{R}$ from the time-time and space-space components of Eq. (2), we obtain moreover the algebraic expression

$$\frac{c^2 K}{R^2} = \frac{8\pi\rho}{3\phi} - H^2 + \frac{\omega}{6}D^2 - HD \tag{7}$$

Finally, by assuming a standard particle content, the state equation is given by the usual form [37] and the energy-momentum conservation law gives the usual expression for the time evolution of the temperature $T$

$$\frac{dt}{dT} = -\frac{d\rho_1/dT}{3H(\rho_1 + P_1/c^2)} \tag{8}$$

where $\rho_1 = \rho_b + \rho_e + \rho_\gamma$, $P_1 = P_e + P_\gamma$ and subscripts $b$, $e$ and $\gamma$ refer to baryon, electron-positron, and photon, respectively.

The differential equation system (Eqs. [6], [8], and the definitions of $H$ and $D$) can be integrated by taking as variable the temperature and an explicit form for the coupling function $\omega(\phi)$ (see Sect. II C). Thus, $R$, $\phi$, $H$, $D$ and $t$ are considered as independent functions.

### B. Boundary Conditions

In order to build up realistic cosmological models, the boundary conditions are fixed, as in Refs. [27–30,35], by the present values of dynamical functions. By present values (hereafter denoted by a sub-index 0) of dynamical functions we mean their current observational values or their limits from observational data.

Some of these values, namely $T_0$ and $H_0$, can be directly measured from observations, whereas $\phi_0$, $D_0$, and $c^2 K/R_0^2$ can be expressed in terms of other better known parameters. As a matter of fact, the $\phi_0$ value can be obtained from $\omega_0$ provided that the coupling function $\omega(\phi)$ has been specified. On the other hand, by using Eqs. (6a) and (7), $D_0$, and $c^2 K/R_0^2$ can be replaced by $q_0 \equiv (-R\ddot{R}/\dot{R}^2)_0 = (-(H^2 + \dot{H})/H^2)_0$ and $\rho_0$:

$$D_0 = \frac{H_0}{2\mathcal{F}_0}\left\{1 \pm \sqrt{1 + 4\mathcal{F}_0\left(q_0 - q_0^{FRW}\frac{3+\omega_0}{2+\omega_0}\right)}\right\} \qquad (9)$$

and

$$\frac{c^2 K}{R_0^2} = \frac{8\pi}{3\phi_0}\rho_0 - H_0^2 + \frac{\omega_0}{6}D_0^2 - H_0 D_0 \qquad (10)$$

where $\mathcal{F}_0 = \frac{\omega_0}{3} - \frac{\omega_0' \Phi_0}{2(3+2\omega_0)}$, $q_0^{FRW} = 4\pi G_0 \rho_0/3H_0^2$, and $\rho_0 = \rho_{R0} + \rho_{b0}$, with $\rho_{R0} = \frac{1}{2}g_{\text{eff}}(T_0)aT_0^4$, $\rho_{b0} = 6.639 \times 10^{-32}(T_0/2.7)^3 \eta_{10}$, $g_{\text{eff}}(T_0)$ is the effective number of relativistic degrees of freedom, and $\eta_{10}$ is the present baryon-to-photon ratio in units of $10^{-10}$.

Note that, in order to have a real $D_0$, the present value of the deacceleration parameter must satisfy

$$q_0 \geq q_0^{FRW}\frac{3+\omega_0}{2+\omega_0} - \frac{1}{4\mathcal{F}_0} \qquad (11)$$

which, when $\omega_0 \gg 1$ and $\mathcal{F}_0 \gg 1$, implies $q_0 \gtrsim q_0^{FRW}$.

The numerical integration of Eqs. (6)–(8) needs to specify the set of parameters $H_0$, $T_0$, $\omega_0$, $q_0$ and $\rho_0$, together with the usual constants (neutron mean-life, number of light neutrino families) appearing in the state equation. A choice for the double-valued $D_0$ solution given by Eq. (9) must be also specified.

In order to avoid an excessive number of free parameters, we have taken the average values of the present photon temperature $T_0 = 2.73 \pm 0.02$ ($2\sigma$) [38], and of the neutron mean-life, $\tau_n = 889 \pm 4$ s. [39]. The present Hubble parameter has been taken to be 50 Km s$^{-1}$Mpc$^{-1}$ [40], and it has been assumed that the number of light neutrino families, $N_\nu$, is three, in accordance with the LEP and SLC results [41]. We will nevertheless discuss in Sect. III B how our results could be modified by the uncertainties in some of these quantities or by observational constraints different from those used here [42].

We have then considered as free parameters the present values of $\omega_0$, $\rho_0$ (or $\eta_{10}$), and $q_0$. Numerical computations show nevertheless that, for scalar-tensor (ST) theories which converge towards GR, the less stringent bounds are obtained when $q_0 \simeq q_0^{FRW}$. A detailed discussion on this point for some particular cases can be found in [30,29,35]. To facilitate the discussion, we will just present here our results with $D_0^{\pm}(q_0,...)$ chosen to give the most conservative constraints on ST theories.

### C. The coupling function $\omega(\phi)$

In a previous paper [36] we showed that a convenient form for the coupling function $\omega(\phi)$ is given by

$$|3 + 2\omega| = (3/\lambda^2)(x^{-\epsilon} + k) \tag{12}$$

where

$$x = \begin{cases} 1 - \Phi & \text{(if } \Phi < 1) \\ \Phi - 1 & \text{(if } \Phi > 1) \end{cases} \tag{13}$$

and $1/2 < \epsilon < 2$ in order to ensure that theories converge at present towards GR. Eq. (12) gives an exact representation for most of the particular scalar-tensor theories proposed in the literature [44] and, in addition, it contains all the possible early behaviors of any theory where $w(\phi)$ is a monotonic, but arbitrary, function of $\phi$ [36].

Our analysis in terms of the $k$ and $\lambda$ parameters revealed the existence of scalar-tensor cosmological models with early expansion rates different from that found in the standard FRW scenarios [36]. This change in the expansion rate of the universe can modify the PNP in different ways. In the next section we will discuss the light element yields by distinguishing four main classes of viable scalar-tensor cosmological models (Table 1). The first and the second class correspond to singular models with a monotonic speed-up factor $\xi(T) \equiv H/H^{FRW}$ faster or slower, respectively, than in GR. Non-singular models or models with a critical temperature where $3 + 2\omega = 0$ constitute the third class. Finally, the last class corresponds to models with a non-monotonic $\xi(T)$ function.

We know that each one of these classes can be subdivided according to other criteria as, for instance, the $\xi$ value at the singularity, the increasing or decreasing evolution of $\Phi$.... Such analysis has been performed in our previous paper [36]. However, we will see that the classification using the four above classes is enough for the purposes of this paper.

### III. RESULTS

#### A. Primordial Nucleosynthesis Bounds

Light element production in scalar-tensor theories has been computed by using the updated reaction rates given by Caughlan and Fowler [45] and Smith et al. [42]. Boundary conditions were set as explained in Sect. II B. We will discuss here the yields of light elements as a function of the baryon-to-photon ratio in units of $10^{-10}$, $\eta_{10}$, the present value of the coupling function, $\omega_0$, and the parameters $k$, and $\lambda$, characterizing the form of $\omega(\Phi)$ (see Eq. [12]). Although we now consider $\epsilon = 1$, we will show in the next subsection that our conclusions just depend very weakly on this parameter. Only the $^4$He, (D/H) and (D + $^3$He)/H primordial yields have been used to constraint each scalar-tensor theory. We have also computed the $^7$Li/H primordial abundance (briefly discussed in section III B) but, due to its well-known uncertainties, it has not been used in our analysis.

*1. Singular models with a monotonic $\xi(T)$ faster than in GR (Class-1)*

Some scalar-tensor models (see Figure 1) imply that, during all the PNP, the expansion rate of the universe is faster than in GR. The range of parameters defining these models can be seen in Table 1. Most of the particular scalar-tensor theories proposed in the literature are included within this category.

Our computations show that all these theories have qualitatively similar implications on the primordial abundance of light elements. Before a general discussion, we first present as an example our results for the theory defined by $\lambda^2 = 3/2$, $k = 0$ and $\epsilon = 1$.

Figures 2 show the primordial abundance of $^4$He (denoted as $Y_p$), (D/H), and (D + $^3$He)/H as a function of $\omega_0$ and $\eta_{10}$. We see from these figures that, for any fixed value of $\eta_{10}$, these abundances increase as $\omega_0$ decreases. This behavior can be interpreted in the following way. When $\omega_0$ decreases, the expansion rate $\xi$ during nucleosynthesis is faster (Figure 1). The temperature $T_*$ at which the n/p ratio freezes out is then higher [1], and this in turn implies a higher light element production [46]. On the other hand, the $\eta_{10}$ dependence of primordial abundances for a given value of $\omega_0$ is such that the $^4$He production increases with $\eta_{10}$ while those of D/H and (D+$^3$He)/H decrease. The reason of this behavior is that reaction rates increase with $\eta_{10}$ and, consequently, the $^4$He production from the D and $^3$He burning is more efficient for larger values of $\eta_{10}$.

In order to analyze the compatibility between observed and predicted light-element yields, we have depicted the observational bounds [42] by a thick line on the abundance axes of Figures 2. As can be seen from these figures, the constraints on this theory are essentially imposed by $^4$He, and (D + $^3$He)/H. To have a right abundance of these elements, the conditions $\omega_0 \gtrsim 10^{21}$ and $\eta_{10} \simeq 3$ are required. In particular, for very large $\omega_0$ values,

---

[1] The freezing out temperature, $T_*$, is roughly determined by the condition: $\lambda(T_*)t(T_*) \simeq 1$, where $\lambda(T_*)$ is the rate for the weak interactions changing protons into neutrons and vice versa, and $t(T_*)$ is the age of the universe.

$^4$He requires $1 \lesssim \eta_{10} \lesssim 3.7$, while $(D + {}^3He)/H$ imposes $\eta_{10} \gtrsim 2.8$. Consequently, the interval $2.8 \lesssim \eta_{10} \lesssim 3.7$ is required to have simultaneously right primordial yields for all the light elements. This $\eta_{10}$ interval is even more narrow for smaller values of $\omega_0$ until, for $\omega_0 \lesssim 10^{21}$ (or, equivalently, $\xi_{10} \equiv \xi(10^{10}K) \lesssim 1.024$), simultaneously right abundances are not further obtained for any value of $\eta_{10}$.

Primordial nucleosynthesis bounds on other class-1 scalar-tensor theories, with different $\lambda$, $k$ and $\epsilon$ parameters, can be obtained by performing a similar computation as before. However, since the predicted abundances in this class of cosmological models essentially depend on the $T_*$ value (or, equivalently, on the $\xi_{10}$ value), the upper bound on $\xi_{10}$ will remain nearly the same whatever the theory is

$$\xi_{10} \lesssim 1.024. \tag{14}$$

On the contrary, since such a $\xi_{10}$ value is obtained for $\omega_0$ values depending on the theory under consideration, the $\omega_0$ bounds will be in general different. The $k$ and $\lambda$ dependence of these $\omega_0$ bounds can be deduced from Figs. 3a and 3b, where we show the results for that value of $\eta_{10}$ (=3) leading to the less restrictive constraints. We see from these figures that the larger $\lambda$ and the smaller $k$, the higher bounds on $\omega_0$ are found. Such a dependence does not however imply small $\omega_0$ bounds in the limits of small $\lambda$ or large $k$ values because, in these limits, the constraint on $\omega_0$ does not further decrease and it remains equal to

$$\omega_0 \gtrsim 10^{21}. \tag{15}$$

Clearly, these bounds on $\omega_0$ are much more stringent than those obtained from the Post-Newtonian experiments ($\omega_0 \gtrsim 500$). They imply that any class-1 scalar-tensor theory is indistinguishable from GR from the beginning of PN up to the present. The allowed range for $\eta_{10}$ is then similar to the usual one ($2.8 \lesssim \eta_{10} \lesssim 3.7$), what in turn implies the usual constraints on the baryonic contribution to the density parameter

$$0.01 \lesssim \Omega_{b0} \lesssim 0.09 \tag{16}$$

where the relation

$$\Omega_{b0}h^2 = 3.77 \times 10^{-3}\eta_{10}(T_0/2.73)^3/f(\Phi_0) \qquad (17)$$

has been used, and where we have adopted the more conservative uncertainty $0.4 \lesssim h \lesssim 1$ [42] for the Hubble parameter (in units of 100 km s$^{-1}$ Mpc$^{-1}$). The correction factor $f(\Phi_0)$ appearing in Eq. (17) takes into account that, in scalar-tensor theories, the density parameter has an additional dependence on $\Phi$. However, since $f(\Phi_0) = \Phi_0[1 - (\omega_0/6)(D_0/H_0)^2 + (D_0/H_0)]$, where $(D_0/H_0) \simeq 1/\omega_0$ and $|\Phi_0| \simeq 1/\omega_0^{1/\epsilon}$ (see Eqs. 7, 9 and 12), the strong PNP bounds on $\omega_0$ (15) imply that $f(\Phi_0)$ is extremely close to unity and, consequently, its influence on $\Omega_{b0}$ is absolutely negligible.

### 2. Singular models with a monotonic $\xi(T)$ slower than in GR (Class-2)

In models where $\xi$ is, during nucleosynthesis, effectively slower than in GR (see Figure 1 and Table 1 for the range of parameters $\lambda$ and $k$ implying this kind of behaviors), the freezing-out temperature $T_*$ is now smaller than in GR. The smaller $\omega_0$, the smaller the $T_*$ value is, and also the light-element production. Such a $\omega_0$ dependence is shown in Fig. 4 for the theory defined by $\lambda^2 = 3/4$, $k = 0$ ($\omega_0 < 0$ and $\Phi_0 > 1$). On the other hand, concerning the $\eta_{10}$ dependence of primordial abundances, we have again that the $^4$He production from the D and $^3$He burning increases for larger and larger values of $\eta_{10}$. If the expansion rate is slower than in GR, this burning works over a longer period. Consequently, the smaller $\xi_{10}$ (or equivalently $\omega_0$ for this class of models), the stronger the D/H and (D+$^3$He)/H decreasing with $\eta_{10}$ is. Compatibility with observations requires again rather stringent bounds on $\omega_0$, $\xi_{10}$ and $\eta_{10}$. When $\omega_0 \gg 10^{21}$ (or $\xi_{10} \simeq 1$), the lower and upper bounds on $\eta_{10}$ are respectively imposed by (D+$^3$He)/H and $^4$He, like in class-1 theories. However, when $\omega_0 \sim 10^{21}$ ($\xi_{10} < 1$), the resulting underproduction of $^4$He can be balanced by increasing $\eta_{10}$. In this last case, the upper bound on $\eta_{10}$ is instead imposed by D/H and, hence, this bound is expected to have a value somewhat larger than in class-1 models.

Primordial nucleosynthesis bounds on other class-2 scalar-tensor theories have been deduced from similar computations. Figure 3c illustrates the case $\eta_{10} = 3$. When $\lambda^2$ is very

close to $1^-$, we get

$$\omega_0 \gtrsim 10^{21}, \tag{18}$$

and the expansion rate of the universe can deviate from that obtained in GR by at most 4%

$$0.96 \lesssim \xi_{10} \leq 1. \tag{19}$$

These bounds are stronger for theories defined by smaller $\lambda$ values (see Fig. 3c), contrary to what we found in models of class-1. A $k$ dependence is now meaningless because, in order to have the kind of behavior defining this class of theories, a $k = 0$ value is required. Finally, the allowed range of $\eta_{10}$ is somewhat wider than that obtained in FRW cosmologies ($2.8 \lesssim \eta_{10} \lesssim 8.4$). This in turn also implies a slightly wider range for $\Omega_{b0}$

$$0.01 \lesssim \Omega_{b0} \lesssim 0.2 \tag{20}$$

### 3. Models with a cut-off: nonsingular or $3 + 2\omega = 0$ at a finite temperature (Class-3)

Nonsingular models have a maximum value of the universe temperature, $T_{max}$, which corresponds to the minimum of the scale factor. Similarly, models where $3 + 2\omega$ vanishes at some finite temperature have also a maximum value of $T$. The smaller $\omega_0$, the smaller the $T_{max}$ value is. Consequently, in this class of models, the bounds on $\xi_{10}$ and $\omega_0$ are mainly imposed by the condition that $T_{max}$ must be high enough to allow the PNP itself. Figures 5 show, as a function of $k$ and $\lambda$, the minimum $\omega_0$ satisfying this last condition. As can be seen from these figures, the occurrence of PN is only possible for very large values of $\omega_0$ implying an expansion rate during nucleosynthesis extremely close to that obtained in FRW models. Thus, in spite of its very different behavior at early times, this class of theories must be extremely close to GR from the beginning of PN up to the present. The required bounds on $\eta_{10}$ and $\Omega_{b0}$ are then very close to the standard ones (see Sect. III A 1).

The $\lambda$ and $k$ dependence of $\omega_{0,min}$ is different for different intervals of these parameters. Theories defined by $\omega_0 > 0, \Phi_0 < 1$ (Fig. 5a) imply larger $\omega_{0,min}$ values for larger $\lambda$ or smaller $k$ values (note that we do not write $|k|$). The same dependence on $\lambda$ but the opposite on $k$ is however found in theories defined by $\omega_0 < 0, \Phi_0 < 1$ (Fig. 5b). Finally, theories with $\omega_0 < 0, \Phi_0 > 1$ (Fig. 5c) imply larger values of $\omega_{0,min}$ for smaller $k$ values, but they have a non-monotonic dependence on $\lambda$. No class-3 models exist for $\omega_0 > 0, \Phi_0 > 1$.

It is important to note that an appropriate choice of $\lambda$ (for instance, the limit $\lambda^2 \to 0$ in Figs. 5a or 5b) and $k$ could allow for not very large values of $\omega_{0,min}$ ($< 10^{21}$). However, in that case, the bounds on $\omega_0$ and $\xi_{10}$ are those found for models with a monotonic $\xi(T)$ (Sections III A 1 and III A 2).

### 4. Singular models with a non-monotonic $\xi(T)$ function (Class-4)

Theories with a non-monotonic evolution of the speed-up factor $\xi$ (see Fig. 1 and Table 1) have an initial phase where the expansion of the universe is slower than in GR but, afterwards, it becomes faster than in the standard cosmology. Figs. 7 show the light element yields as a function of $\omega_0$ and $\eta_{10}$ in the theory defined by $\lambda^2 = 3/2$ and $k = 1/2$. For extremely large $\omega_0$ values, the maximum of $\xi$ is reached much before the PNP (Fig. 1). Consequently the speed-up factor is greater than unity during nucleosynthesis and the resulting bounds on $\omega_0$, $\xi_{10}$ and $\eta_{10}$ are then equal to those found for theories of class-1 (Sect. III A 1). However, we can also see from these figures that each element has another allowed interval for not so large values of $\omega_0$. In this last interval, simultaneously right primordial abundances can be obtained for large $\eta_{10}$ values. For instance, if $\omega_0 = 6 \times 10^{17}$, $^4$He requires $10.0 \lesssim \eta_{10} \lesssim 15.2$; D/H requires $\eta_{10} \lesssim 11.6$ and (D+$^3$He)/H imposes $\eta_{10} \gtrsim 4$, what leaves the allowed range $10.0 \lesssim \eta_{10} \lesssim 11.6$. Other values of $\omega_0$ in this particular theory, imply different bounds on $\eta_{10}$ making possible any value between 2.8 and 12.4. The last upper bound on $\eta_{10}$ needs a present value of the coupling function $\omega_0 \gtrsim 10^{15}$ and, hence, the theory can differ from GR much more than in the previous classes ($\xi_{max} \lesssim 1.6$).

The above example shows that the right primordial abundances can be obtained in the framework of a class-4 theory clearly distinguishable from GR. Furthermore, this class of theories allows for quite large $\eta_{10}$ values. The achievement of such $\eta_{10}$ values can be explained in the following way. In the previous particular theory, the expansion rate of the universe at the beginning of nucleosynthesis is slower than in GR ($\xi_{10} = 0.72$ if $\omega_0 = 6 \times 10^{17}$ and $\eta_{10} = 11$). Similar to what we found for Class-2 theories, there exists a tendency to the underproduction of $^4$He, which can be balanced by considering larger $\eta_{10}$ values. Consequently, the upper bound on $\eta_{10}$ is here constrained by the D/H abundance. However, unlike class-2 theories, the expansion rate of the universe becomes during the PNP faster than in GR. The D burning is not then very effective because it occurs in a shorter time and, hence, large $\eta_{10}$ values are allowed.

In order to obtain the largest $\eta_{10}$ value allowed in the framework of this class of models, we must then analyze those theories with the largest $\xi$ value during the PNP, but with $\xi_{10} < 1$ at the beginning of PN. These two conditions are better obtained if the slope of $\xi(T)$, in its increasing interval, is as high as possible. Figures 8 and 9 show the $\lambda$ and $k$ dependence of primordial abundances in this class of theories with $\eta_{10} = 3$. We see that the form of these curves is similar to that of $\xi(T)$, (especially $Y_p$ because it has the strongest dependence on $\xi_{10}$). Consequently, any discussion about the behavior of $\xi(T)$ will be performed in terms of, for example, $Y_p(\omega_0)$. The sharpness of the maximum of $\xi$ depends mainly on $k$. The smaller $k$, the sharper the maximum of $\xi$ is. On the other hand, the $\xi_{max}$ value increases as $\lambda$ decreases while the $k$ dependence of $\xi_{max}$ is non-monotonic, with higher $\xi_{max}$ values in the limits of very large and very small $k$ values. In order to have a very large $\xi$ value during nucleosynthesis we then need to analyze the limit of very small $\lambda$ and $k$ values. We have performed such an analysis and we have found that, in fact, the allowed $\Omega_{b0}$ value is considerably increased with respect to that obtained in other classes of scalar-tensor theories. However, the lower bound on $\omega_0$ is smaller as the allowed $\Omega_{b0}$ is larger. The upper bound on the baryon contribution to the density parameter is then determined by the compatibility with solar-system experiments, which requires $\omega_0 > 500$. This is found for the theory $\lambda^2 = 0.08$, and $k = 2 \times 10^{-7}$, which leads to right primordial abundances with $\eta_{10} \leq 58.7$, that is

$$0.01 \lesssim \Omega_{b0} \lesssim 1.38 \tag{21}$$

The above result does not signify that $\Omega_{b0}$ values equal or greater than unity are only possible for rather low values of $\omega_0$. For instance, for $\lambda^2 = 0.1$ and $k = 10^{-5}$, (see Fig. 10), we can in fact always find some interval of $\omega_0$ greater than $10^7$ allowing nevertheless for $\eta_{10}$ values in the range $2.8 \lesssim \eta_{10} \lesssim 56.5$. The resulting permitted interval for $\Omega_{b0}$ is

$$0.01 \lesssim \Omega_{b0} \lesssim 1.33 \tag{22}$$

Closure of the universe by baryons is then possible in the framework of this class of scalar-tensor theories even when a standard composition of the universe is used [47]. The $\omega_0$ values needed to allow for $\Omega_{b0} \geq 1$ are very large and, consequently, these of theories are at present close enough to GR to ensure compatibility with post-Newtonian experiments. On the contrary, they are very different from GR at early times and imply expansion rates of the universe which can be several hundred of times faster (when $\xi = \xi_{max}$) than in the FRW cosmology.

### B. Uncertainties on the Primordial Nucleosynthesis bounds

#### 1. Influence of parameters

In the above analysis we have taken $N_\nu = 3$, $\tau_n = 889$ s, $T_0 = 2.73$, $H_0 = 50$, and $\epsilon = 1$ to avoid an excessive number of free parameters. We will now analyze how the uncertainties on these parameters could modify our bounds on ST theories.

The width of the $Z^0$ particle is directly related to the number of light particles which can couple to the $Z^0$. The measurements of the $Z^0$ width from results of the LEP and SLC $e^+e^-$ colliders [41] imply that the number of light neutrino species is $N_\nu \leq 3.25$ at the $2\sigma$ level. Thus, the choice of $N_\nu$ equal to the three known species of light neutrinos does not seem at present a source of uncertainty in PN results [48].

Determinations of the neutron half-life based on the storage of ultra-cold neutrons [39] have considerably reduced the uncertainty in this parameter. By combining these with other recent results, Smith et al. [42] infer $\tau_n = 889.8 \pm 3.6$ s. We have performed some computations where the neutron mean-life was varied within the above allowed range. We obtained that, when $\tau_n$ is changed from 886.2 to 893.4 s, primordial abundances vary as $\Delta Y_p = +0.0015$, and $\Delta \text{D/H} = \Delta(\text{D}+^3\text{He})/\text{H} = +0.018 \times 10^{-5}$. This implies an uncertainty of $\pm 0.01$ for our bounds on $\xi_{10}$ and possible differences in a factor of two in our constraints on $|\omega_0|$. The uncertainty in the neutron mean-life has then a very small effect on our conclusions of section III A implying that theories with a monotonic evolution of $\xi$ are not significantly different from GR. The constraints on $\omega_0$ remain very strong ($|\omega_0| \gtrsim 10^{20}$) and the deviation from GR at the beginning of PN is at most 3.4% in class-1 theories, or 5% in class-2 theories. In the same way, our results concerning models with a non-monotonic $\xi(T)$ do not change significantly. This class of theories can deviate strongly from GR and allows for very large $\eta_{10}$ values. The uncertainty in $\tau_n$ can only modify the bounds on $\eta_{10}$ given in section III A 4 by $\pm 0.3$.

The uncertainty in the other usual PN parameters ($T_0$ and $H_0$) has no significant effect on primordial abundances for a given $\xi_{10}$ value. The Hubble parameter can however produce an uncertainty in the $\omega_0$ value leading to that speed-up factor, $\xi_{10}$, at the beginning of PN. For instance, computations with $h = 1$ lead to bounds on $|\omega_0|$ which are smaller by a factor of 10 than those obtained with $h = 0.5$. The constraints on $\xi_{10}$ do not change and neither those on $\eta_{10}$. As before, the high $\omega_0$ bounds we have found imply that the uncertainty introduced on this constraint is not physically important.

With respect to the $\epsilon$ influence, our computations with $1/2 < \epsilon < 2$ show that the resulting bounds on $\omega_0$ are slightly smaller if $\epsilon$ is greater than unity (they decrease at most by a factor of $\simeq 10$ when $\epsilon$ is close to 2), but they increase very strongly for $\epsilon$ values smaller than unity (for example, when $\epsilon = 0.95$ in a class-1 theory, the bound on $\omega_0$ typically increase by a factor of $10^6$, then implying $\omega_0 \gtrsim 10^{27}$). On the contrary, the bounds on $\xi_{10}$ and $\eta_{10}$ do not depend significantly on $\epsilon$. We then find that our constraints on scalar-tensor theories cannot be relaxed by the influence of this parameter more than by considering, for example, the effect of the uncertainty in $H_0$.

## 2. Consequences due to other observational constraints on primordial abundances

Concerning the observational constraints on primordial abundances, we have used in this paper the most widely accepted ones. Recent measures of primordial abundance of deuterium [49] from the absorption spectra of high-redshift quasars seem however indicate a rather high observational lower bound for D/H ($\simeq 1.9 \times 10^{-4}$). These measures are at present preliminary and cannot be reconciled with the upper bound of (D+$^3$He)/H ($\simeq 9 \times 10^{-5}$). If we however accept such a constraint for D/H and we then ignore that for (D+$^3$He)/H, the upper bound $\eta_{10}$ is considerably reduced in GR and also in scalar-tensor theories. In particular, in GR and scalar-tensor theories with a monotonic $\xi(T)$, compatibility with such observations would require $\eta_{10} \lesssim 1.6$ and $\Omega_{b0} \lesssim 0.04$, which leaves few room for non-luminous baryonic matter in the Universe. This effect would be however much less dramatic for scalar-tensor theories with a non-monotonic $\xi(T)$. For example, in the theory defined by $\lambda^2 = 0.1$ and $k = 10^{-5}$ (see Eq. 22), simultaneously right abundances of $^4$He and D/H can be obtained for $\eta_{10} \lesssim 12.5$ and $\omega_0 \gtrsim 5 \times 10^7$, which implies $\Omega_{b0} \lesssim 0.3$. This last bound can be increased to $\Omega_{b0} \lesssim 0.4$ by considering other class-4 theories. These values of $\Omega_{b0}$ are compatible with most of measures of the total density parameter $\Omega_{b0}$. Consequently, even when the high-redshift quasar measures are considered, class-4 theories allow for large $\eta_{10}$ values and a universe content largely dominated by baryons.

Finally, we remark that primordial productions of $^7$Li/H were computed in all runs of our nucleosynthesis code. However, due to its well known uncertainties, we have not used that element to constraint scalar-tensor theories. We note however that all the above constraints imply a $^7$Li/H production compatible with the more conservative observational constraint for this element ($^7$Li/H$< 1.3 \times 10^{-9}$). Consequently, our bounds would not be modified by considering this abundance. This is also true for class-1 to class-3 theories when a more severe observational range is used ($1.0 \times 10^{-10}$ $<^7$Li/H$< 2.3 \times 10^{-10}$). However, the $\Omega_{b0}$ upper-bound obtained in class-4 theories would be reduced by a factor of $\sim 0.5$, then implying $\Omega_{b0} \lesssim 0.65$.

## IV. CONCLUSIONS AND GENERAL DISCUSSION

Primordial production of light elements has been calculated in the framework of Scalar-Tensor cosmological models for a sweep of initial conditions compatible with astronomical data. Our results and conclusions are absolutely different for theories with a monotonic evolution of the speed-up factor $\xi(T)$ and those with a non-monotonic $\xi(T)$.

In the first case, our results imply that the expansion rate of the universe at the beginning of primordial nucleosynthesis can differ from that obtained in the usual FRW model by at most 4%

$$0.96 \lesssim \xi_{10} \lesssim 1.024 \tag{23}$$

Consequently, the present value of the coupling function must be greater than

$$\omega_0 \gtrsim 10^{20} \tag{24}$$

and, if $\xi > 1$, the allowed range for the baryon density parameter is essentially the same as in GR

$$\begin{aligned} 2.8 &\lesssim \eta_{10} \lesssim 3.7 \\ 0.01 &\lesssim \Omega_{b0} \lesssim 0.09 \end{aligned} \quad \text{(class-1)} \tag{25}$$

or, if $\xi < 1$, just slightly wider

$$\begin{aligned} 2.8 &\lesssim \eta_{10} \lesssim 8.4 \\ 0.01 &\lesssim \Omega_{b0} \lesssim 0.2 \end{aligned} \quad \text{(class-2)} \tag{26}$$

These bounds, together with the cosmological evolution of Scalar-Tensor theories studied in paper I [36], imply that all these cosmological models are indistinguishable from the standard FRW ones from the beginning of PN up to the present (except for slight differences on the $\eta_{10}$ bounds in class-2 models). The uncertainties on the usual input PN parameters can just slightly relax the above bounds and, hence, they do not modify our conclusions.

Primordial nucleosynthesis is then a very strong test for the viability of this kind of gravitational theories. Very small deviations from GR during the early stages of the evolution of the universe imply a light element production inconsistent with present observations. However, we cannot assure that these theories are also indistinguishable from GR at very early epochs before nucleosynthesis or with a non-standard composition of the universe [36,50].

We would like also to note that the general PN constraints given by Eq. (24) for Scalar-Tensor theories with a monotonic $\xi(T)$ are of the same order of magnitude as those obtained by us and by other authors in the framework of the particular cases proposed in the literature [29–35]. These PN constraints are instead much higher than those implied by the post-Newtonian experiments [21], and also than those found by Damour and Nordtvedt [51] (who estimated the matter-dominated evolution of $\omega(\phi)$ in flat scalar-tensor theories by assuming $\omega(\phi) \simeq 1$ at the beginning of that era). Primordial Nucleosynthesis then arises as a test which imposes constraints on the early value of $\omega(\phi)$ which are roughly similar to those found from post-Newtonian experiments for the present $\omega_0$ value. However, since $\omega(\phi)$ increases during the universe evolution, such constraints are much stronger at the beginning of the matter-dominated era, and even much stronger at present. Note also that, by construction, other cosmological tests (measured values of $H_0$, $q_0$, $T_0$, etc. [40]) are also satisfied in all the models analyzed in this paper.

Concerning theories with a non-monotonic $\xi(T)$, our analysis leads to very different conclusions. This class of theories can reproduce the right yields of light elements even when they are very different from GR during the radiation-dominated era. In some theories, the constraint on $\omega_0$ is similar to that imposed by solar system experiments

$$\omega_0 \gtrsim 500. \tag{27}$$

In such cases, the maximum value of $\xi$ is much higher than unity

$$\xi_{max} \lesssim 350 \tag{28}$$

and the allowed range for the baryon density parameter is much wider than in GR

$$2.8 \lesssim \eta_{10} \lesssim 58.7$$

$$0.01 \lesssim \Omega_{b0} \lesssim 1.38 \qquad (29)$$

Allowed $\Omega_{b0}$ values greater than unity can also be obtained in other class-4 theories implying a much larger present value of the coupling function $\omega_0$. For instance, theories as that of Eq. (22), imply $\omega_0 \gtrsim 10^7$ while the upper limit of $\Omega_{b0}$ is only slightly smaller than that given by Eq. (29): $\Omega_{b0} \lesssim 1.33$. The universe closure by baryons is then possible in the framework of these theories without requiring a non-standard composition for the cosmic gas [47].

Furthermore, we have shown in Sect III B that, if we accept some recent measures of D/H from high-redshift quasars, the resulting upper bound on $\Omega_{b0}$ would be drastically reduced in the framework of GR as well as in theories with a monotonic $\xi(T)$: $\Omega_{b0} \lesssim 0.04$. On the contrary, the $\Omega_{b0}$ bounds in models with a non-monotonic $\xi(T)$ would be much less severely affected by such measures. They would be still much larger ($\Omega_{b0} \lesssim 0.4$) than the usual one. On the other hand, it is also important to note that all the bounds given by the above expressions also imply a right abundance of $^7$Li/H when the most conservative observational constraints ($^7$Li/H $< 1.3 \times 10^{-9}$) are considered. Nevertheless, if we use more severe observational constraints ($1.0 \times 10^{-10} \lesssim {}^7$Li/H $\lesssim 2.3 \times 10^{-10}$), the upper bound on $\Omega_{b0}$ imposed by the class-4 theories would be $\Omega_{b0} \lesssim 0.65$, which is still considerably wider than the usual one.

This is the first time that a class of scalar tensor theories is found to be compatible both with primordial abundances and with solar-system experiments while remaining clearly distinguishable from GR. This last feature is not only manifested in the much wider allowed range for $\eta_{10}$ but also in the cosmological evolution of the universe during the PNP and, in the extreme case of Eq. (21), during all the radiation dominated era.

### ACKNOWLEDGMENTS

We are grateful to L. Blanchet for valuable comments about this work.

# REFERENCES

[1] D. La and P. J. Steinhardt, Phys. Rev. Lett. **62**, 376 (1989).

[2] A. R. Liddle and D. Wands, Phys. Rev. D **45**, 2665 (1992).

[3] P. Jordan, Nature **164**, 637 (1949).

[4] E. W. Kolb, M. J. Perry, and T. P. Walker, Phys. Rev. D **33**, 869 (1986).

[5] E. Vayonakis, Phys. Lett. B **213**, 419 (1988).

[6] A. A. Coley, Astron. Astrophys. **233**, 305 (1990).

[7] P. S. Wesson, Astrophys. J. **394**, 19 (1992).

[8] Y. M. Cho, Phys. Rev. Lett. **68**, 3133 (1992)

[9] P. S. Wesson and J. Ponce de Leon, Astron. Astrophys. **294**, 1 (1995).

[10] M. M. Green, J. H. Schwarz and E. Witten, Superstring theory (Cambridge Univ. Press, Cambridge, 1988)

[11] C. Brans and R. H. Dicke, Phys. Rev. **124**, 925 (1961).

[12] P. A. Dirac, *Physicist's Conception of Nature*, ed. J. Mehra (Dordrecht, Reidel, 1973).

[13] B. M. Barker, Astrophys. J. **219**, 5 (1978).

[14] J. D. Bekenstein, Phys. Rev. D 15, 1458 (1977).

[15] G. Schmidt, W. Greiner, U. Heinz and B. Muller, Phys. Rev. D **24**, 1484 (1981).

[16] P. G. Bergmann, Int J. Theor. Phys. **1**, 25 (1968).

[17] R. V. Wagoner, Phys. Rev. D **1**, 3209 (1970).

[18] K. Nordtvedt, Astrophys. J. **161**, 1059 (1970).

[19] V. M. Canuto, P. J. Adams, S. H. Hsieh and E. Tsiang, Phys. Rev. D **16**, 1643 (1977)

[20] V. M. Canuto and I. Goldman, Nature **296**, 709 (1982).

[21] C. M. Will; Phys. Reports **113**, 345 (1984).

[22] R. V. Wagoner, Astrophys. J. **179**, 343.

[23] J. D. Barrow, Month. Not. Royal Astron. Soc. **184**, 677.

[24] J. Yang, D. N. Schramm, G. Steigman and R. T. Rood, Astrophys. J. **277**, 697.

[25] F. S. Acceta, L. M. Krauss and P. Romanelli, Phys. Lett. B **248**, 146.

[26] J. A. Casas, J. García-Bellido and M. Quirós, Phys. Lett. B **278**, 94.

[27] G. S. Greenstein, Astrophys. Space Sci. **2**, 155.

[28] K. Arai, M. Hashimoto and T. Fukui, Astron. Astrophys. **179**, 17.

[29] G. Yepes and R. Domínguez-Tenreiro, Phys. Rev. D **34**, 3584 (1986); R. Domínguez-Tenreiro and G. Yepes, Astron. Astrophys. **177**, 5 (1987).

[30] A. Serna, R. Domínguez-Tenreiro and G. Yepes, Astrophys. J. **391**, 433 (1992).

[31] J. D. Bekenstein and A. Meisels, Astrophys. J. **237**, 342 (1980); A. Meisels, Astrophys. J. **252**, 403 (1982).

[32] T. Rothman and R. Matzner, Astrophys. J. **257**, 450 (1982)

[33] V. M. Canuto and I. Goldman, Nature **304**, 311 (1983).

[34] R. Domínguez-Tenreiro and A. Serna, Gen. Relativ. Grav. **22**, 1207 (1990); A. Serna and R. Domínguez-Tenreiro, Astrophys. J. **389**, 1 (1992).

[35] A. Serna and R. Domínguez-Tenreiro, Phys. Rev. D **47**, 2363 (1993); Phys. Rev. D **48**, 1591 (1993).

[36] A. Serna and J. M. Alimi, Phys. Rev. D. (submitted).

[37] R.A. Alpher, J.W. Follin Jr and R.C. Herman, Phys. Rev. **92**, 1347 (1953).

[38] H. P. Gush, M. Halpern and E. Wishnow, Phys. Rev. Lett. **65**, 537 (1990); A. Kogut et al., Astrophys. J. **355**, 102 (1990); J. C. Mather et al. Astrophys. J. Letters **354**, L37 (1990); E. Palazzi et al., Astrophys. J. **357**, 14 (1990); G. De Amicci, Astrophys. J. **381**, 341 (1991).

[39] G. P. Yost et al., Phys. Lett. B **204**, 1 (1988); W. Mampe, P. Ageron, C. Bates, J. M. Pendlebury and A. Steyerl, Phys. Rev. Lett. **63**, 593 (1989); V. P. Afimenkov et al., JETP Lett. **52**, 373 (1990).

[40] G. de Vaucouleurs and H. G. Cowin, Jr., Astrophys. J. **297**, 23 (1985); W. D. Arnett, D. Branch and J. C. Wheeler, Nature **314**, 337 (1985); A. Sandage, Astrophys. J. **331**, 583 (1988); A. Sandage and G. A. Tamman, Astrophys. J. **365**, 1 (1990); A. Sandage et al., Astrophys. J. **401**, L7 (1992).

[41] P. Aarnio et al. (DELPHI collaboration), Phys. Lett. B **231**, 539 (1989); G. I. Abrams et al., Phys. Rev. Lett. **63**, 724 (1989); B. Adeva et al. (L3 collaboration), Phys. Lett. B **231**, 509 (1989); B. Adeva et al. (L3 collaboration), Phys. Lett. B **275**, 209 (1992); M. Z. Akrawy et al. (OPAL collaboration), Phys. Lett. B **231**, 530 (1989); D. Decamp et al. (ALEPH collaboration), Phys. Lett. B **231**, 519 (1989); D. Decamp et al. (ALEPH collaboration), Phys. Lett. B **235**, 399 (1990); J. M. Dorfan et al. (MARK II collaboration), Phys. Rev. Lett. **63**, 2173 (1989).

[42] M. S. Smith, L. H. Kawano and R. A. Malaney, Astrophys. J. Suppl. **85**, 219 (1993).

[43] J. García-Bellido and M. Quirós, Phys. Lett. B **243**, 45 (1990).

[44] In fact, Brans-Dicke's theory is found by setting $\epsilon = 0$ in Eq. (12; Barker's theory corresponds to $\lambda^2 = 3$, $k = 0$, and $\epsilon = 1$; Schmidt-Greiner-Heinz-Muller's theory corresponds to $k = \lambda^2 + 1$ and $\epsilon = 1$. Finally, General Relativity can be obtained by taking $\Phi = 1$.

[45] G. Caughlan and W. A. Fowler, Atomic Data Nuclear Data Tables **40**, 291 (1988).

[46] A. M. Boesgaard and G. Steigman, Ann. Rev. Astron. Astrophys. **23**, 319 (1985).

[47] J.-M. Alimi and A. Serna, Nature (submitted).

[48] R. A. Malaney and G. J. Mathews, Phys. Reports **229**, 145 (1993).

[49] J. K. Weeb, R. F. Carswell, M. J. Irwin, and M. V. Penston, Mon. Not. Roy. Astron. Soc. **250**, 657; A. Songaila and L. L. Cowie, Nature (submitted 1994).

[50] R. Domínguez-Tenreiro and G. Yepes, Astrophys. J. (Letters) **317**, L1 (1987); G. Yepes and R. Domínguez-Tenreiro, Astrophys. J. **335**, 3 (1988).

[51] T. Damour and K. Nordtvedt, Phys. Rev. D **48**, 3436 (1993).

| $\omega_0 > 0$ | | | | | | | $\omega_0 < 0$ | | | | |
|---|---|---|---|---|---|---|---|---|---|---|---|
| $\Phi_0 > 1$ | | $\Phi_0 < 1$ | | | | | $\Phi_0 > 1$ | | | | $\Phi_0 < 1$ |
| $k > 0$ | | $k \leq 0$ | $k > -1$ | | | $k \leq -1$ | $k \neq 0$ | $k = 0$ | | | |
| $\frac{\lambda}{\sqrt{k}} > 1$ | $\frac{\lambda}{\sqrt{k}} \leq 1$ | | $0 < \frac{\lambda}{\sqrt{k+1}} < 1$ | $1 < \frac{\lambda}{\sqrt{k+1}} < 3$ | $k = 0, \lambda = 1$ | | | $\lambda > 1$ | $\lambda = 1$ | $\lambda < 1$ | |
| Class-4 | Class-1 | Class-1 | Class-1 | Class-3 | GR-like | Class-3 | Class-3 | Class-1 | GR-like | Class-2 | Class-3 |

Table 1: Range of parameters defining the four main classes of scalar-tensor cosmological models

FIGURES

FIG. 1. Expansion rate $\xi = H/H^{FRW}$ as a function of $T$ for different types of cosmological models. Solid lines correspond to $\omega_0$ values smaller than those of dashed-lines.

FIG. 2. Primordial abundances of a) $^4$He, b) D/H, and c) (D+$^3$He)/H, as a function of $\omega_0$. The theory shown in this figure is defined by $\omega_0 > 0, \Phi_0 > 1, k = 0, \lambda^2 = 3/2$. Solid line corresponds to $\eta_{10} = 1$, dashed line to $\eta_{10} = 3$, and dotted line to $\eta_{10} = 5$.

FIG. 3. $\lambda$, $k$ and $\omega_0$ dependence of $Y_p$ in theories with a monotonic $\xi(T)$. Theories shown in this figure are defined by a) $\omega_0 > 0, \Phi_0 > 1, k = 3$ and $\lambda^2 = 1.5$ (solid lines), 0.75 (dashed lines), and 0.01 (dotted lines); b) $\omega_0 > 0, \Phi_0 > 1, \lambda^2 = 3/2$ and $k = 26.5$ (solid lines), 10 (dashed lines), and 3 (dotted lines); and c) $\omega_0 < 0, \Phi_0 > 1, k = 0$ and $\lambda^2 = 0.75$ (solid lines), 0.5 (dashed lines), and 0.3 (dotted lines).

FIG. 4. Primordial abundances of a) $^4$He, b) D/H, and c) (D+$^3$He)/H, as a function of $\omega_0$. The theory shown in this figure is defined by $\omega_0 < 0, \Phi_0 > 1, k = 0, \lambda^2 = 3/4$. Solid line corresponds to $\eta_{10} = 1$, dashed line to $\eta_{10} = 3$, and dotted line to $\eta_{10} = 5$.

FIG. 5. $\lambda$ dependence of the smallest value of $\omega_0$ needed to have a high enough temperature for the PNP in theories with a cut-off. a) $\omega_0 > 0, \Phi_0 < 1$, and $k = 0.5$ (solid line), -2.5 (dashed line), and -23.5 (dotted line). b) $\omega_0 < 0, \Phi_0 < 1$, and $k = 10$ (solid line), 1 (dashed line), and -8 (dotted line). c) $\omega_0 < 0, \Phi_0 > 1$, and $k = -0.5$ (solid line), -3 (dashed line), and -5 (dotted line).

FIG. 6. Primordial abundances of a) $^4$He, b) D/H, and c) (D+$^3$He)/H, as a function of $\omega_0$. The theory shown in this figure is defined by $\omega_0 > 0, \Phi_0 > 1, k = 1/2, \lambda^2 = 3/2$. Solid line corresponds to $\eta_{10} = 3$, dashed line to $\eta_{10} = 5$, and dotted line to $\eta_{10} = 10$.

FIG. 7. $\lambda$, $k$ and $\omega_0$ dependence of $Y_p$ in theories with a non-monotonic $\xi(T)$. Theories shown in this figure are defined by a) $\omega_0 > 0, \Phi_0 > 1, k = 0.01$ and $\lambda^2 = 1.$ (solid lines), 0.4 (dashed lines), and 0.3 (dotted lines); and b) $\omega_0 > 0, \Phi_0 > 1, \lambda^2 = 3/2$ and $k = 1.$ (solid lines), 1/2 (dashed lines), and 0.01 (dotted lines).

FIG. 8. Primordial abundances of a) $^4$He, b) D/H, and c) (D+$^3$He)/H, as a function of $\eta_{10}$. The theory shown in this figure is defined by $\omega_0 > 0, \Phi_0 > 1, \lambda^2 = 10^{-1}$ and $k = 10^{-5}$. Lines correspond to $\omega_0 = 10^7$ (solid lines), $5 \times 10^7$ (dashed lines), and $10^{25}$ (dotted lines). The last case gives cosmological models close to the FRW ones

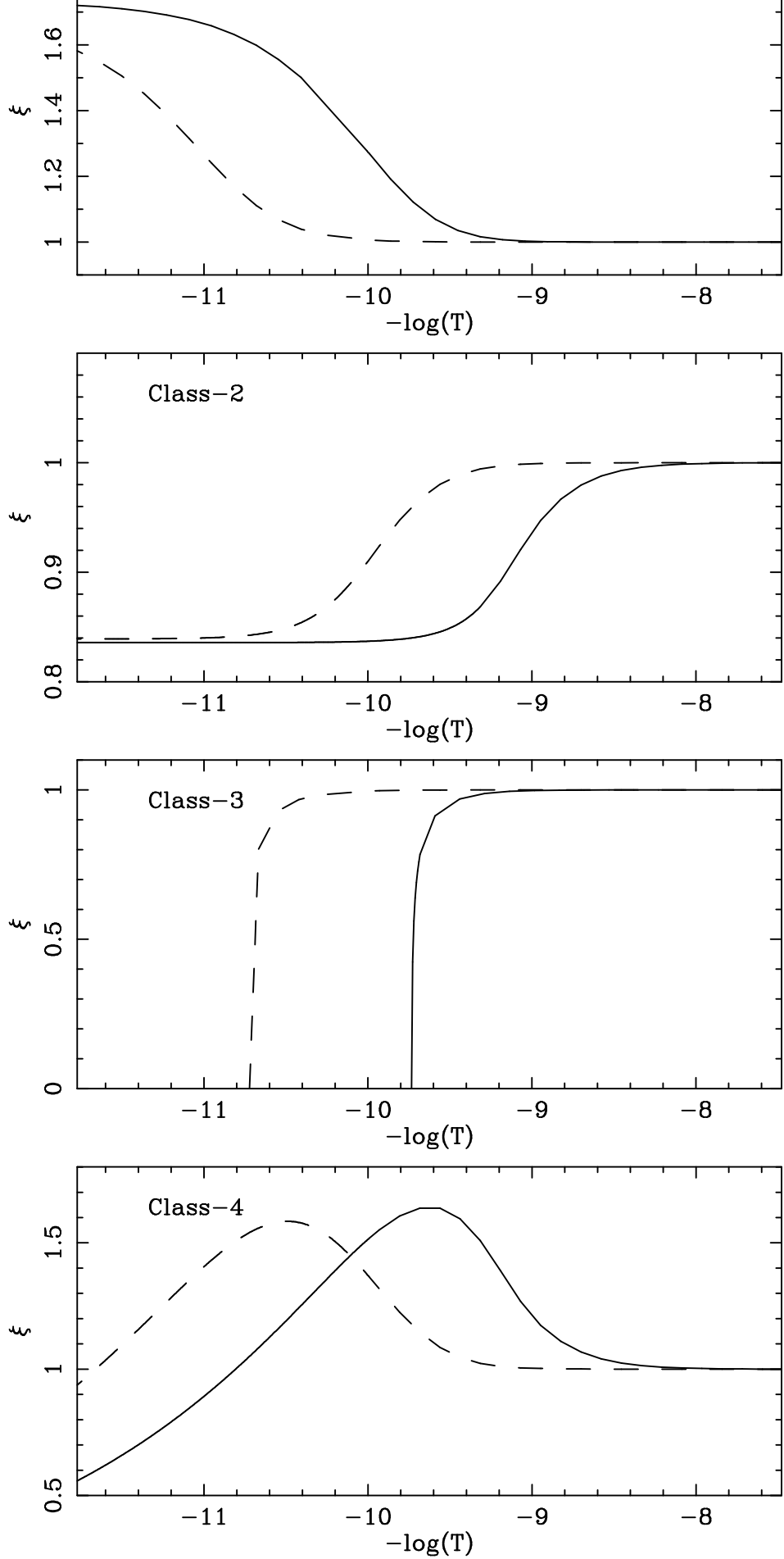

Fig.1

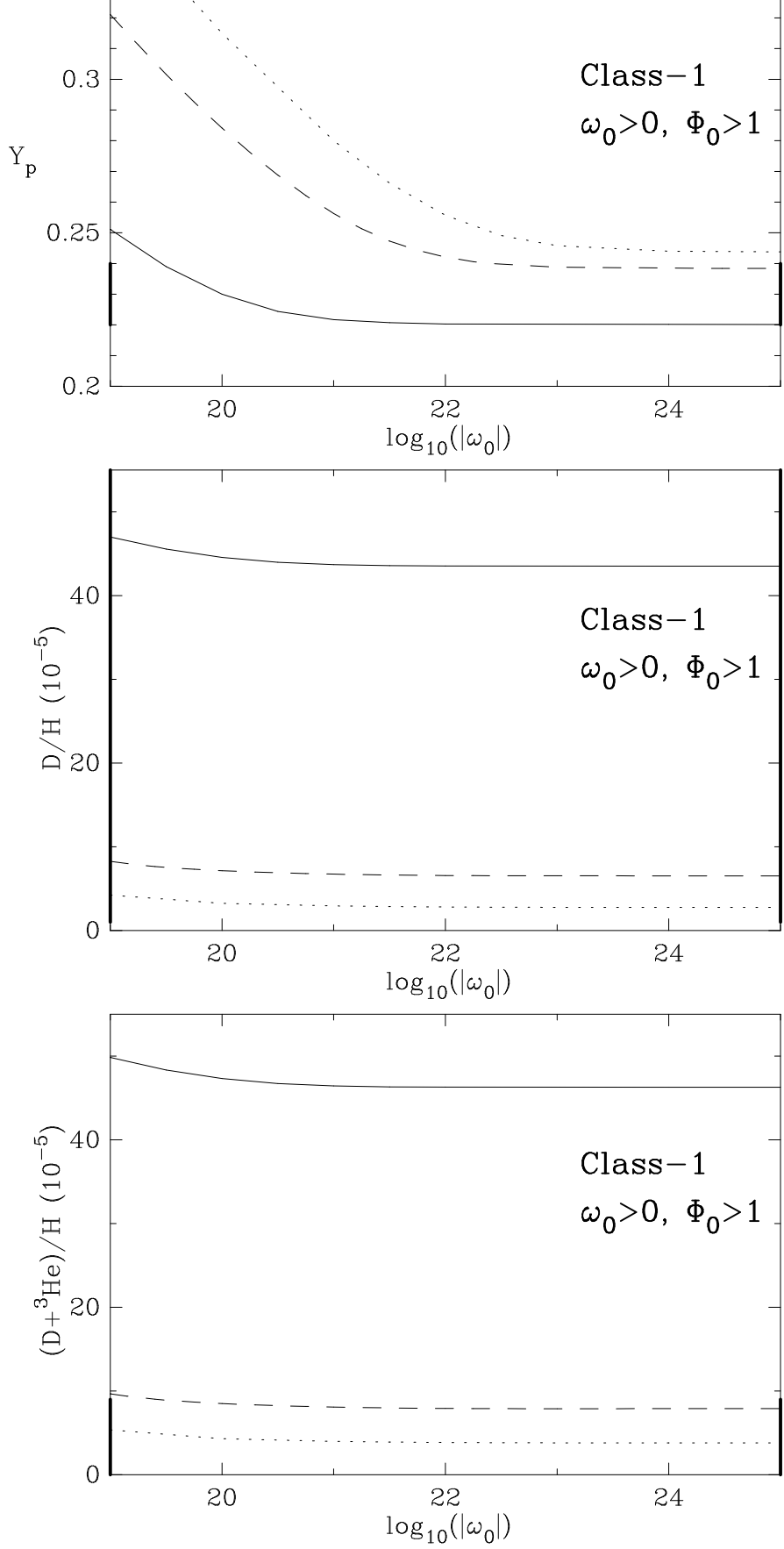

Fig.2

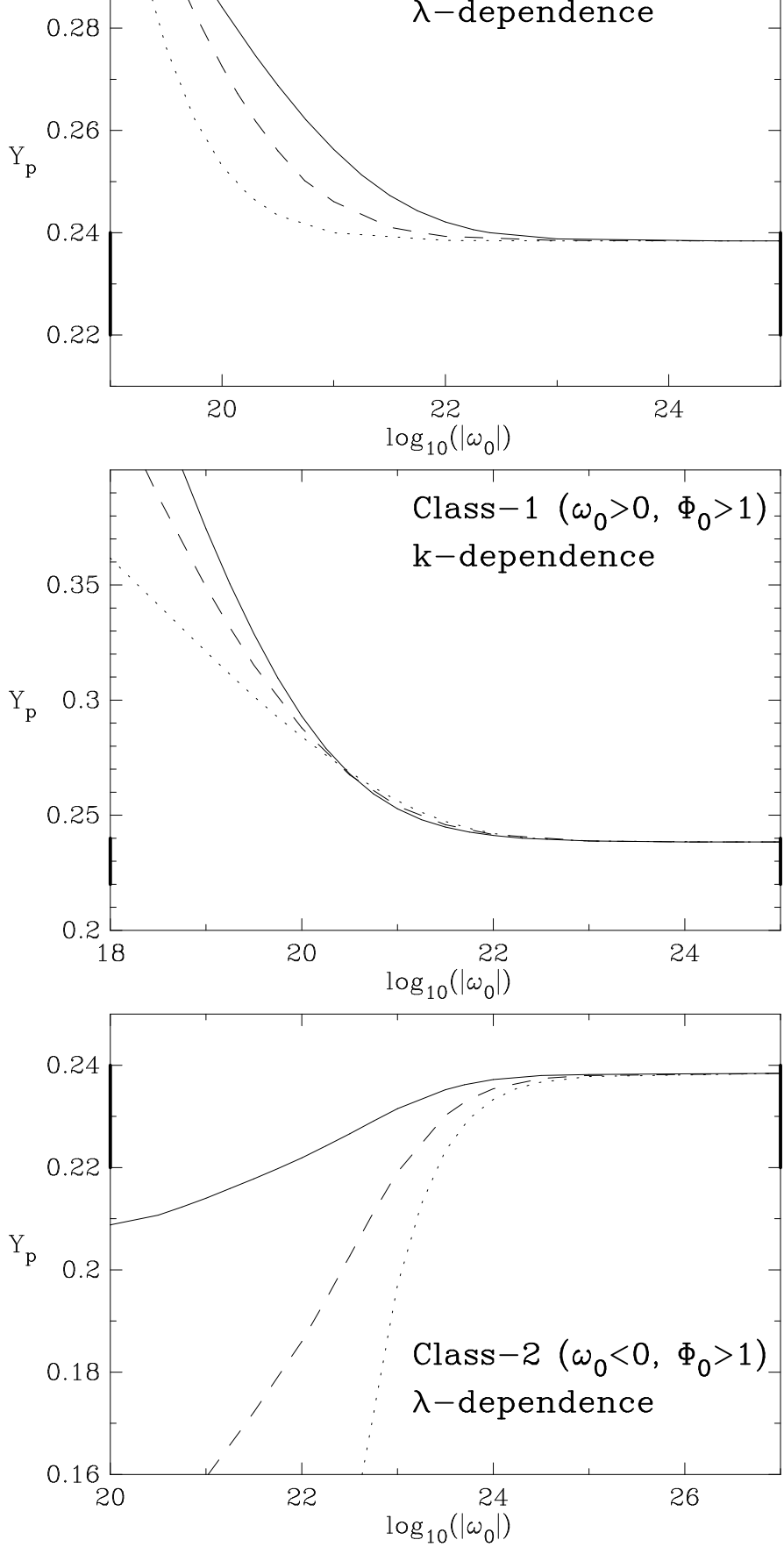

Fig.3

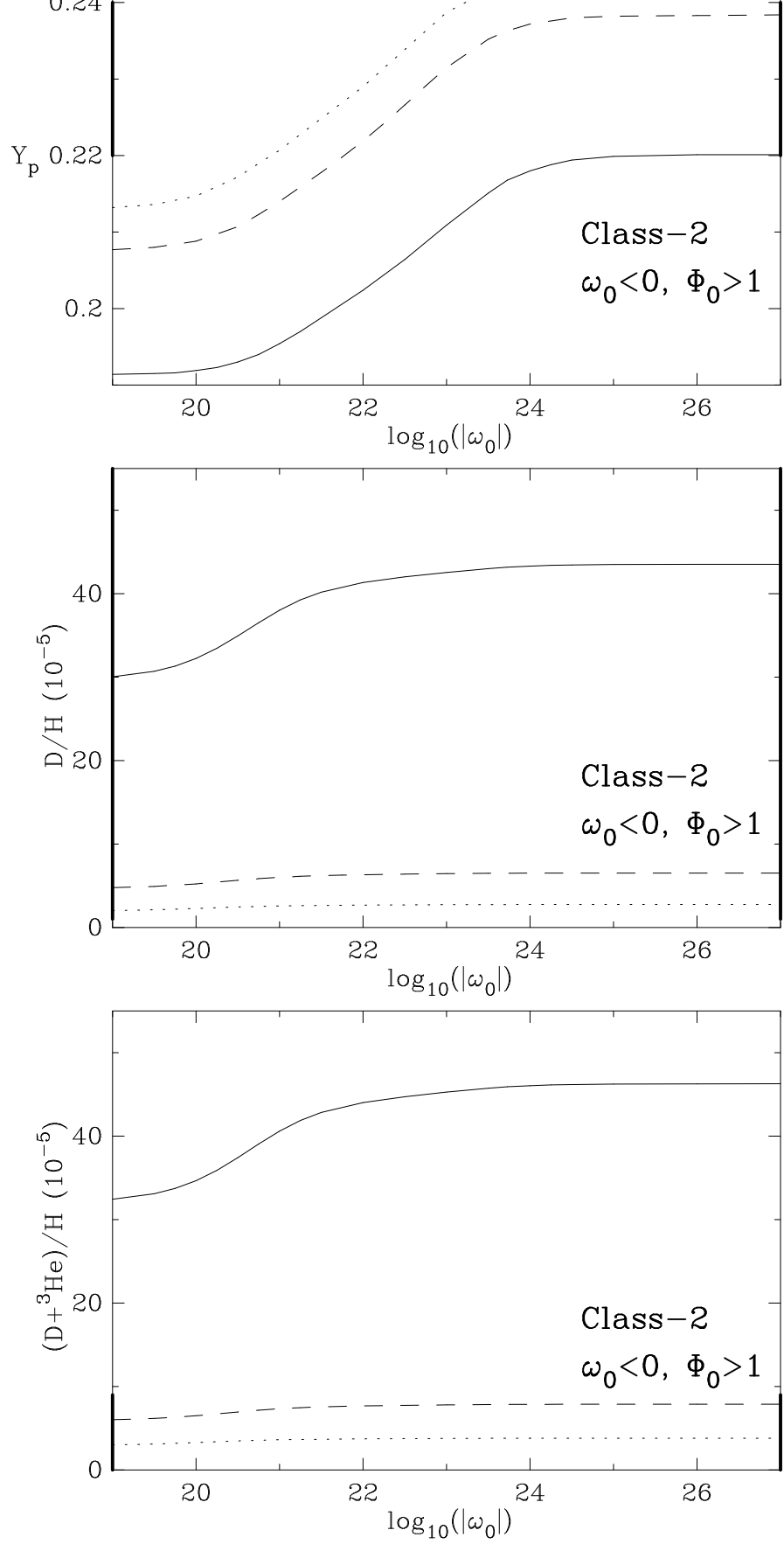

Fig.4

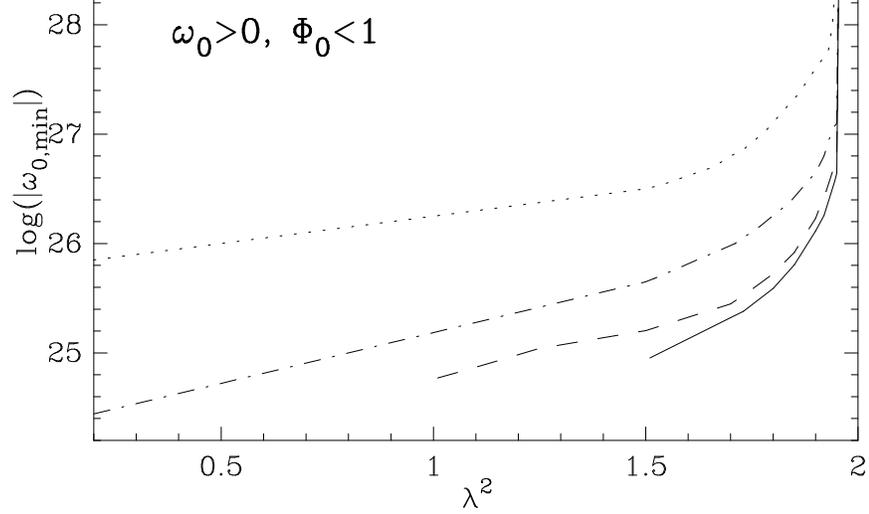
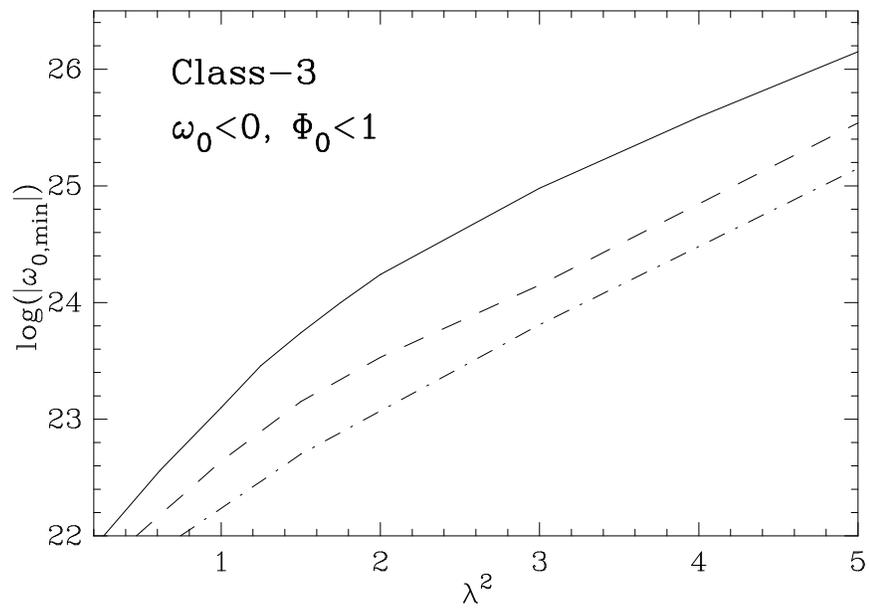
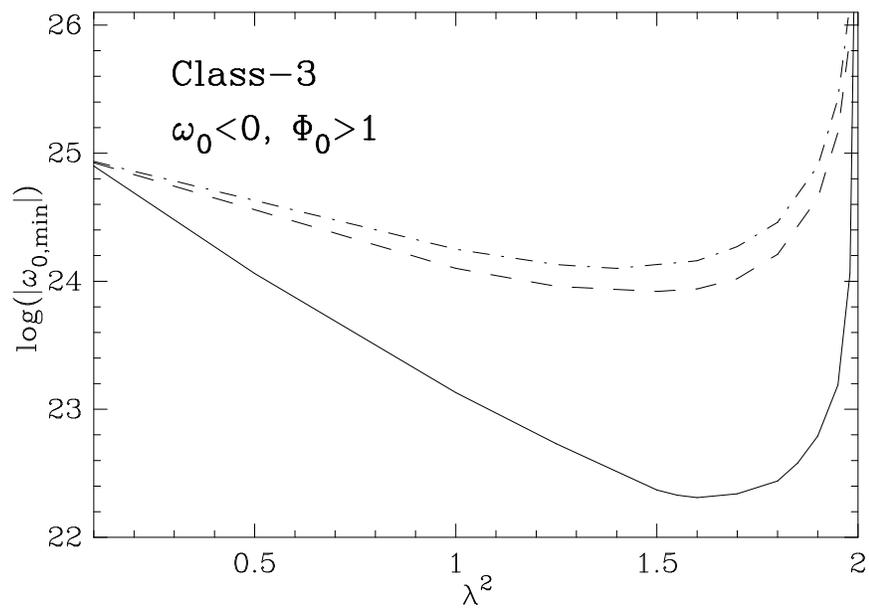

Fig.5

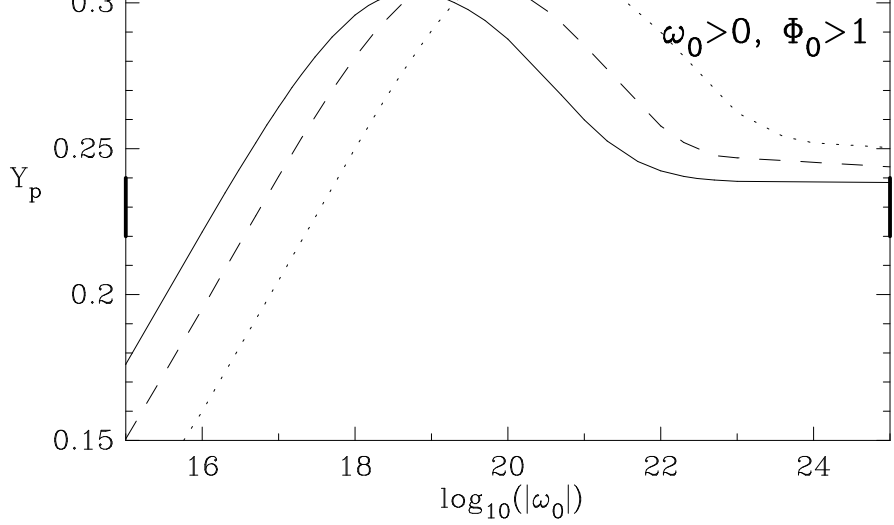
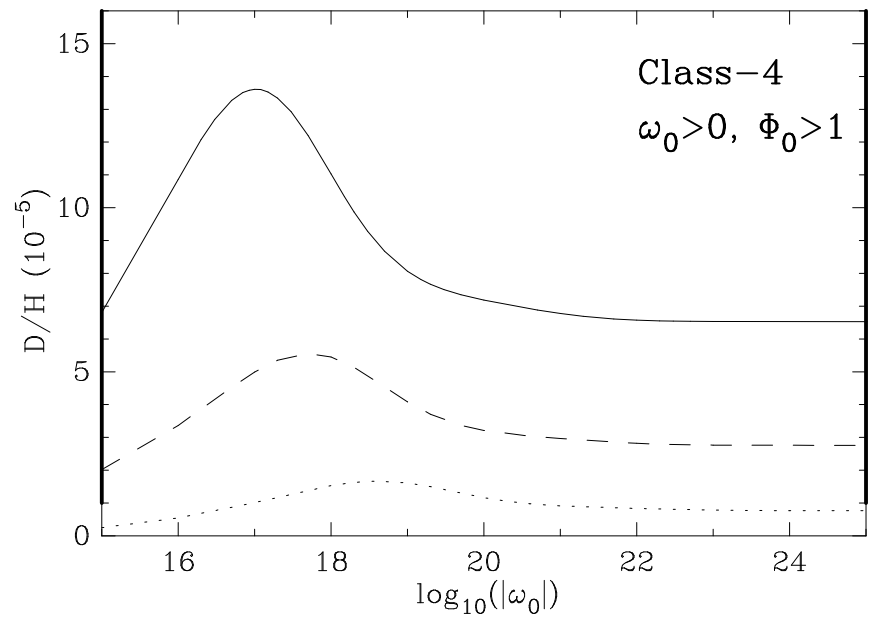
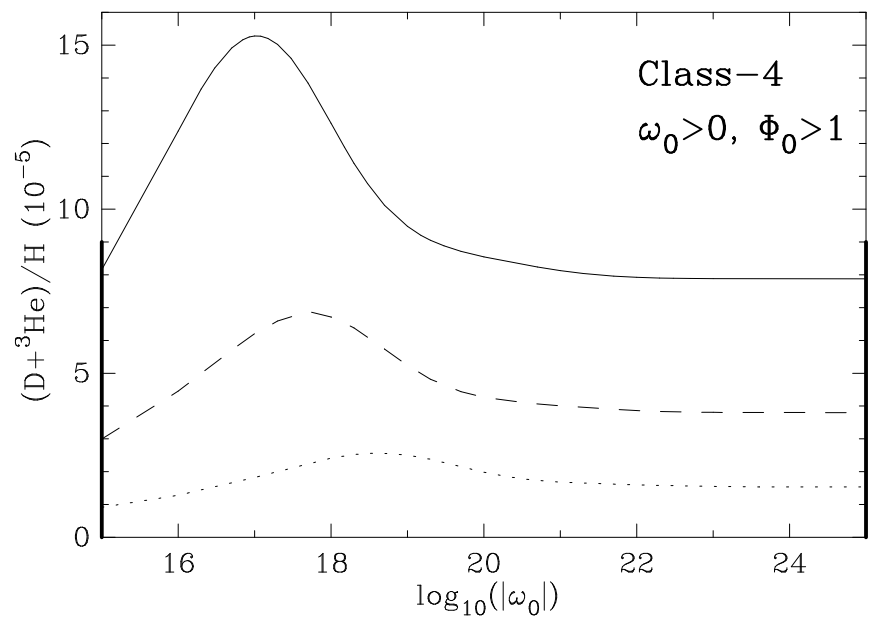

Fig.6

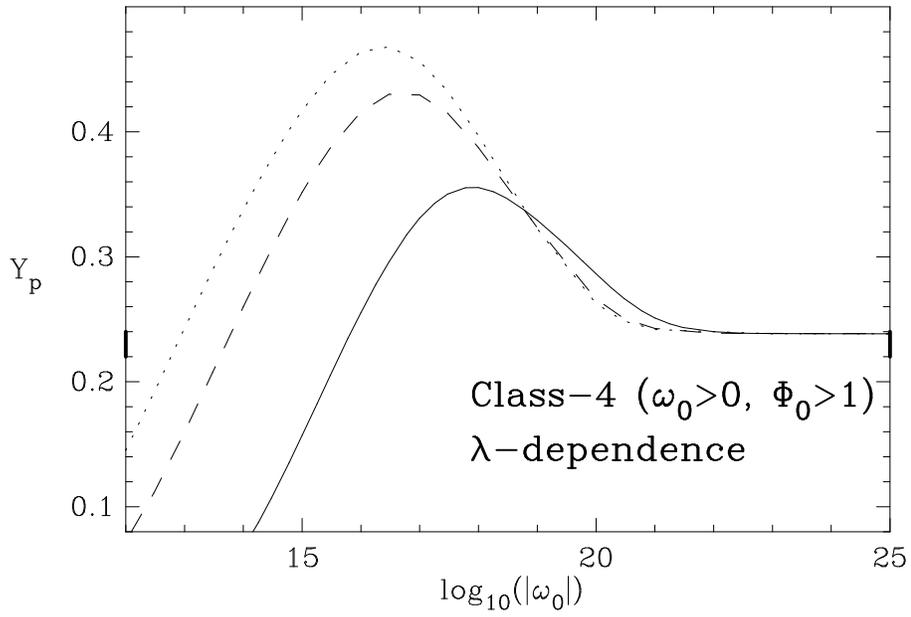

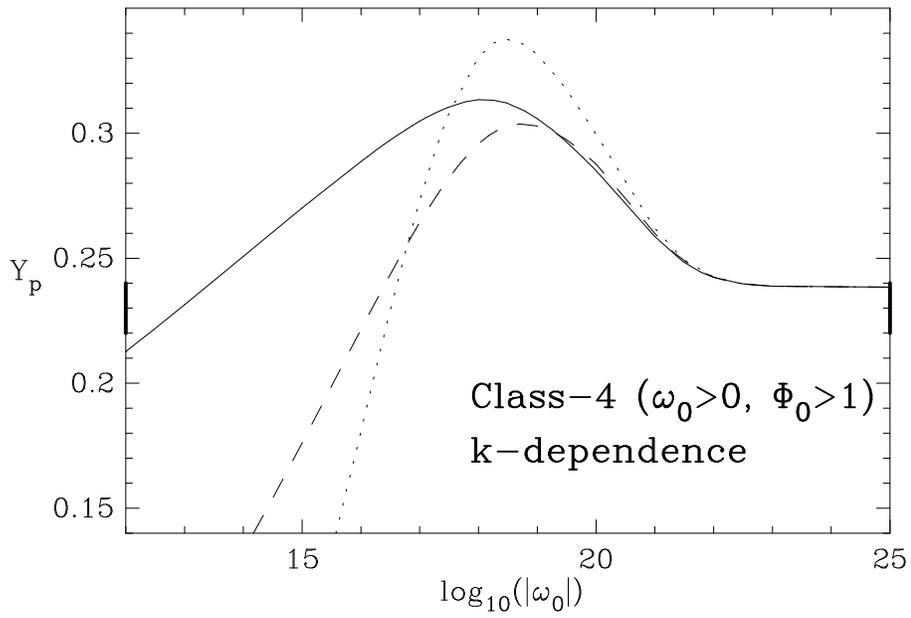

Fig.7

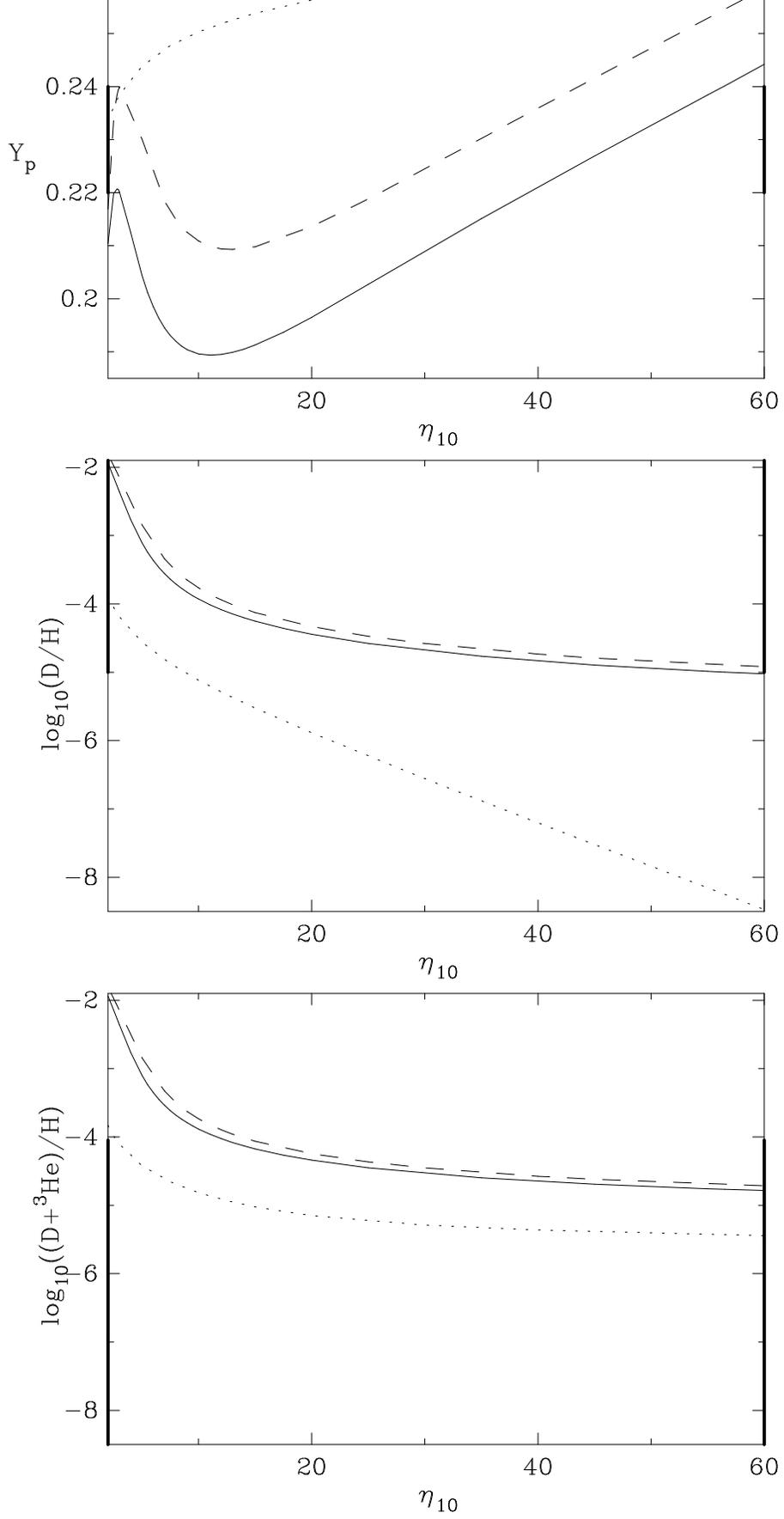

Fig.8